\documentclass[aps,prl,preprint,psfig,groupedaddress]{revtex4-1}

\usepackage{amsmath,amssymb}
\usepackage{graphicx}
\usepackage{bm}
\usepackage{epstopdf} 

\AppendGraphicsExtensions{.tif}

\newcommand{\BS}{Bi$_2$Se$_3$}
\newcommand{\BST}{(Bi,Sb)$_2$Te$_3$}

\newcommand{\LIR}{$\lambda_{\rm{IREE}}$}
\newcommand{\YIG}{Y$_3$Fe$_5$O$_{12}$}
\newcommand{\vsp}{$V_{\rm{SP}}$}
\newcommand{\gsm}{$g_{\uparrow \downarrow}$}
\renewcommand{\vec}[1]{\mbox{\boldmath$1$}}

\newcounter{lastnote}

\def\bc{\begin{center}}
\def\ec{\end{center}}
\def\be{\begin{equation}}
\def\ee{\end{equation}}
\renewcommand{\vec}[1]{\mbox{\boldmath$1$}}

\begin{document}
\title{Surface state dominated spin-charge current conversion in topological insulator/ferromagnetic insulator heterostructures}
\author{Hailong Wang$^1$, James Kally,$^1$, Joon Sue Lee,$^1$ Tao Liu,$^2$ Houchen Chang,$^2$ Danielle Reifsnyder Hickey,$^3$ Andre Mkhoyan,$^3$ Mingzhong Wu,$^2$ Anthony Richardella,$^1$ Nitin Samarth$^1$}
\email{nsamarth@psu.edu}
\affiliation{$^1$Department of Physics, The Pennsylvania State University, University Park, Pennsylvania 16802, USA}
\affiliation{$^2$Department of Physics, Colorado State University, Fort Collins, CO 80523, USA}
\affiliation{$^3$Department of Chemical Engineering and Materials Science, University of Minnesota, Minneapolis, Minnesota 55455, USA}

\date{\today}
\begin{abstract}
We report the observation of ferromagnetic resonance-driven spin pumping signals at room temperature in three-dimensional topological insulator thin films -- \BS~ and \BST~ -- deposited by molecular beam epitaxy on \YIG~ thin films. By systematically varying the \BS~film thickness, we show that the spin-charge conversion efficiency, characterized by the inverse Rashba-Edelstein effect length (\LIR), increases dramatically as the film thickness is increased from 2 quintuple layers, saturating above 6 quintuple layers. This suggests a dominant role of surface states in spin and charge interconversion in topological insulator/ferromagnet heterostructures. Our conclusion is further corroborated by studying a series of \YIG/\BST~heterostructures. Finally, we use the ferromagnetic resonance linewidth broadening and the inverse Rashba-Edelstein signals to determine the effective interfacial spin mixing conductance and \LIR.
\end{abstract}
\maketitle

The development of next-generation spintronic devices has driven extensive studies of spin-to-charge conversion through measurements of the inverse spin Hall effect (ISHE) and/or the inverse Rashba-Edelstein effect (IREE) in both three-dimensional (3D) \cite{Kajiwara_2010,Heinrich_2011,Mosendz_2010,Du_2013,Hahn_2013,Wang_2014} and two-dimensional (2D) material systems \cite{Sanchez_2013,Shiomi_2014,Mellnik_2014,Fan_2014,Deorani_2014,Jamali_2015,Baker_2015,Wang_2015,Kondou_arxiv}. Topological insulators (TIs) such as the Bi-chalcogenides are naturally relevant in this context due to the large spin-orbit coupling (SOC) strength and the inherent spin-momentum ``locking" in their surface states \cite{Mellnik_2014,Hsieh_2009,Hasan_RMP} which promise very efficient spin-charge conversion efficiency. Previous studies of spin transfer in TI-based heterostructures have involved ferromagnetic metals that provide a shunting current path, therefore introducing potential artifacts which complicate the picture and analysis \cite{Shiomi_2014,Mellnik_2014,Deorani_2014,Jamali_2015}. To circumvent these problems, we have grown and characterized bilayers of TIs on ferrimagnetic insulator \YIG~(YIG) thin films with an exceptionally low damping constant \cite{Chang_2014}. Here, we report the ferromagnetic resonance (FMR)-driven spin pumping observed in YIG/\BS~bilayers, showing robust spin pumping signals at room temperature. Systematic variation of the \BS~thickness allows us to unambiguously demonstrate that the spin-charge conversion efficiency, characterized by the inverse Rashba-Edelstein effect (IREE) length \LIR~in a 2D material system\cite{Sanchez_2013}, dramatically increases from $(1.1 \pm 0.13)$ pm to $35 \pm 4$ pm as the \BS~thickness varies from 2 to 6 quintuple layers (QL). When the top and bottom surface states with opposite spin polarizations decouple from each other, \LIR~ saturates and is constant, providing clear evidence for the dominant role of surface states in inducing spin-charge conversion in 3D TIs.

We first discuss the structural and interfacial characterization of the YIG/\BS~heterostructure using high-resolution scanning transmission electron microscopy (HR-STEM). Figure 1(a) shows an atomically ordered 6 QL \BS~layer grown on an epitaxial 30-nm YIG thin film. We note that an amorphous layer of about 1 nm in thickness is observed at the YIG/\BS~interface, most likely due to the nucleation of the template layer in the two-step growth process (see Supplementary Material at [link to be added] for more details about the growth method). The atomic force microscopy (AFM) image in Fig. 1(b) shows a smooth surface with a roughness of about 0.71 nm. A representative $\theta -2 \theta$  x-ray diffraction (XRD) scan of a 40 QL \BS~film shown in Fig. 1(c) indicates a phase-pure \BS~layer. Figure 1(d) shows a representative FMR derivative absorption spectrum for a 30-nm YIG film used in this study taken at a radio-frequency (rf) $f = 3$ GHz with a magnetic field $H$ applied in the film plane. The peak-to-peak line width ($\Delta H_{PP}$) obtained from the spectrum is 9.2 Oe, and an effective saturation induction of 1.76 kOe is extracted from fitting the frequency dependence of the resonance field \cite{Chang_2014}.
The spin pumping measurements are performed using a microwave transmission line on the YIG/TI bilayers at room temperature (approximate sample dimensions of 1 mm $\times$  5 mm). During the measurements, a DC magnetic field $H$ is applied in the $x$-$z$-plane and the spin pumping voltage \vsp~is measured across the $\sim 5$ mm long TI layer along the $y$-axis, as illustrated in Fig. 2(a). At the resonance condition, the YIG magnetization $M$ precesses around the equilibrium position and transfers angular momentum to the conduction electrons in the TI films through interfacial exchange coupling \cite{Du_2013}. The resulting pure spin current  is injected along the $z$-axis with spin polarization $\sigma$ parallel to $M$, and then converted to a charge current leading to the spin pumping signals.

Figure 2(b) shows the temperature dependence of the resistivity of 6 QL \BS~and \BST~thin films grown on YIG. The metallic behavior (decrease in resistivity at low temperature) is the typical behavior of \BS~due to Se vacancies \cite{Kandala_2013}. For \BST, the resistivity increases by $\sim 50 \%$ from room temperature to 2 K, consistent with surface state dominated transport in this thin film \cite{JSLee_2015}. The carrier concentrations obtained from Hall effect measurements at room temperature are $4 \times 10^{13}$ cm$^{-2}$ and $9.8 \times 10^{12}$ cm$^{-2}$ for 6 QL \BS~ and \BST, respectively.

Figure 2(c) shows the observed \vsp~vs. $H$ spectra of the YIG/\BS~(6QL) bilayers at $f =$ 2, 3, and 4 GHz using 100 mW microwave power. The observed spin pumping signals change sign when the magnetic field $H $ is reversed from $\theta_H = 90^{\circ}$ to $270^{\circ}$, as expected from either IREE or ISHE. At 2 and 3 GHz, the observed signal is about 40 $\mu$V, and for 4 GHz, the signal decreases to about 20 $\mu$V, which results from the variation of the microwave transmission line performance at different frequencies. Figure 2(d) shows the spin pumping spectra of a YIG/\BS~(6 QL) sample at microwave powers of 18, 32, 56, and 100 mW and an excitation frequency of 3 GHz. The upper inset shows the rf-power dependence of \vsp~at $\theta_H = 90^{\circ}$,   indicating that the observed spin pumping signals are in the linear regime.

To probe the spin to charge conversion mechanism in TI layers, we systematically vary the \BS~thickness from 2 to 60 QL. Figure 3(a) shows the spin pumping spectra when $\theta_H = 90^{\circ}$ for 4, 6, 24, and 40 QL thicknesses of \BS~grown on YIG, respectively. The significant enhancement of the spin pumping signal in the low \BS~thickness regime mainly results from the increased resistivity. For a 2D material system, such as the TI surface states, the spin to charge conversion is dominated by IREE [22, 23] and the injected spin current  is converted into a 2D charge current, $J_c =  \lambda_{\rm{IREE}} J_s$. The spin current density $J_s$ is in units of A m$^{-2}$, and the 2D charge current density $J_c$  is in units of A m$^{-1}$; the parameter \LIR~has the dimension of length and is introduced to characterize the spin to charge conversion efficiency in 2D material systems \cite{Sanchez_2013,Shen_2014}. The observed spin pumping voltages \vsp~dominated by IREE depend on several material parameters \cite{Sanchez_2013}:
\begin{equation}
V_{SP}= - w R \lambda_{\rm{IREE}} J_s,
\end{equation}
where $w$ and $R$ are the sample width and resistance, respectively. $J_s$ is the spin current density at the YIG/TI interface which can be expressed as \cite{Mosendz_2010,Hahn_2013,Wang_2014}:
\begin{equation}
{J_{\rm{s}}} = \frac{{2e}}{\hbar }\frac{{{g_{ \uparrow  \downarrow }}{h_{{\rm{rf}}}}^2\hbar {\omega ^2}\left[ {\gamma 4\pi {M_s} + \sqrt {{{\left( {\gamma 4\pi {M_s}} \right)}^2} + 4{\omega ^2}} } \right]}}{{2\pi {{\left( {\Delta {H_{{\rm{pp}}}}} \right)}^2}\left[ {{{\left( {\gamma 4\pi {M_s}} \right)}^2} + 4{\omega ^2}} \right]}},
\end{equation}
where \gsm~is the effective interfacial spin mixing conductance \cite{Tserkovnyak_2005},  $\Delta H_{\rm{pp}}$ is the FMR peak-to-peak linewidth,  $h_{\rm{rf}}$ is the radio frequency field, $\omega$ is the FMR angular frequency, and $M_s$ is the saturation induction of the YIG thin films. We can determine the effective spin mixing conductance \gsm~from the FMR linewidth broadening of the YIG thin film \cite{Heinrich_2011,Mosendz_2010,Tserkovnyak_2005}:
\begin{equation}
g_{ \uparrow  \downarrow } = \frac{{2\pi \sqrt 3 {M_{\rm{s}}}\gamma {t_{{\rm{YIG}}}}}}{{g{\mu _{\rm{B}}}\omega}}\left( {{\Delta H_{{\rm{YIG}}/{\rm{TI}}}} - {\Delta H_{{\rm{YIG}}}}} \right), 
\end{equation} 
where $\gamma$ is the absolute gyromagnetic ratio, $t_{\rm{YIG}}$ denotes the thickness of the YIG thin films, $g$ is the Land\'e factor, and $\mu_B$ is the Bohr magnetron.

If the spin pumping signal is dominated by the ISHE, spin diffusion should be taken into account according to $J_c = {\theta _{{\rm{SH}}}}{\lambda _{{\rm{SD}}}}\tanh \left( {\frac{{{t_{{\rm{TI}}}}}}{{2{\lambda _{{\rm{SD}}}}}}} \right){J_s}$ , and the spin pumping signal will follow \cite{Heinrich_2011,Mosendz_2010,Wang_2014}:
\begin{equation}
{V_{{\rm{SP}}}} =  - wR{\theta _{{\rm{SH}}}}{\lambda _{{\rm{SD}}}}\tanh \left( {\frac{{{t_{{\rm{TI}}}}}}{{2{\lambda _{{\rm{SD}}}}}}} \right){J_{\rm{s}}},
\end{equation}
where $\lambda_{\rm{SD}}$ is the spin diffusion length, $t_{\rm{TI}}$ is the thickness of the TI thin film and $\theta _{\rm{SH}} $ is the spin Hall angle. The distinct difference between Eqs. (1) and (4) is whether the observed spin pumping signal is dominated by the spin momentum ``locking" in the surface states \cite{Liu_2015,Tang_2014,Li_2014} or by the SOC interaction.

To answer this question, Fig. 3(b) shows the \BS~thickness dependence of \vsp~(blue points) and \LIR~(or $J_c / J_s$ )  (red points), where we define  $J_c =  \frac{{{V_{{\rm{SP}}}}}}{{wR}}$. Above 6 QL,  $J_c / J_s$ almost follows a constant value of about 35 pm. Below 6 QL, $J_c / J_s$ dramatically decays by a factor of 30 from $35 \pm 4$ pm  to $1.1 \pm 0.13$ pm when at 2 QL thickness. 
Earlier studies have reported that the thickness of the \BS~surface states is approximately 2-3 nm \cite{YZhang_2010,Neupane_2014}. Above 6 QL, the top and bottom \BS~surface states decouple from each other; below 6 QL, the interaction of the two surface states with opposite spin polarizations can decrease the interfacial spin momentum ``locking" efficiency. This is consistent with angle-resolved photoemission spectroscopy (ARPES) studies that show the opening of a gap in the Dirac cone when the \BS~thickness is below 6 QL, accompanied by a decrease in the spin polarization of the surface states \cite{YZhang_2010,Neupane_2014}. Qualitatively, our data shown in Fig. 3(b) follow this trend and strongly indicate the key role played by the surfaces states in spin-charge conversion in \BS. If we try to interpret the data in Fig. 3(b) with the spin diffusion model (Eq. 4), the fit yields a value of $\lambda_{\rm{SD}} \sim 1.6$ nm and also requires the presence of a ``dead" layer at the interface (see Supplementary Material at [link to be added] for detailed analysis using the spin diffusion model). This short vertical spin diffusion length suggests that the spin polarized electron current is restricted to the bottom surface of the TI. Thus, while we cannot definitively rule out the spin diffusion model, a more physically meaningful picture at this stage is that the surface states probably play a dominant role in the spin-charge conversion. We note that the value we obtain for $( J_c / J_s )$ (or \LIR) is approximately two orders of magnitude smaller than the spin Hall angle reported using a spin torque FMR study at room temperature \cite{Mellnik_2014}. One possible reason for this discrepancy is the amorphous layer at the interface shown in the HR-STEM figure, which potentially decreases the spin injection efficiency. Another reason may be the difference in the fundamental measurement mechanism between these two probing techniques. In a spin torque FMR experiment, as the charge current flows through the TI layers, the electrons can potentially have multiple scattering processes to transfer the spins to the ferromagnetic layers. However, in an FMR spin pumping measurement, this multiple scattering process may not be valid.

To further verify that the spin-charge conversion efficiency is dominated by the surface states of TIs, we grew five different TI heterostructures on YIG as control samples and measured their spin pumping signals. The five control samples are sample A: YIG/\BST~(6 QL); sample B: YIG/\BS~(1 QL)/\BST~(6 QL); sample C: YIG/\BS~(6 QL)/\BST~(6 QL); sample D: YIG/\BS~(1 QL)/Cr$_{0.2}$(Bi$_{0.5}$Sb$_{0.5}$)$_{1.8}$Te$_3$~(6 QL); and sample E: YIG/\BS~(6 QL)/Cr$_{0.2}$(Bi$_{0.5}$Sb$_{0.5}$)$_{1.8}$Te$_3$~(6 QL). Figure 3(c) shows the spin pumping spectra of control samples B, C, D and E at 3 GHz radio-frequency and 100 mW power. The enhancement of the spin pumping signal of samples D and E mainly results from the larger resistivity of Cr$_{0.2}$(Bi$_{0.5}$Sb$_{0.5}$)$_{1.8}$Te$_3$  compared to \BST. Normalizing by the resistance and sample width, we obtained the spin charge conversion ratio of the five control samples and compared them with the values for YIG/\BS~ in Fig. 3(d).
First, the values of \LIR~obtained for sample C and sample E are $37 \pm 4$ pm and $34 \pm 4$ pm, respectively. Both the values are quite close to $35 \pm 4$ pm measured for YIG/\BS~(6QL), indicating that as long as the \BS~thickness is above 6 QL, the spin-charge conversion efficiency is roughly constant and does not depend on the bulk properties: Cr doping and different band structures do not change the  values. Second, for sample A, \BST~directly grown on YIG,  \LIR $= 17 \pm 2$ pm, about half of the value of \BS. This is in sharp contrast with earlier results which reported a much larger spin Hall angle of the \BST~compared with \BS~using a spin-polarized tunneling study \cite{Shen_2014}. This most likely results from the different interfacial quality and conditions that determine the spin momentum ``locking" efficiency. We expect that the bottom surface state condition at the  YIG/\BST~interface \cite{YZhang_2010} is not as good as the CoFeB/MgO/\BST~interface \cite{Shen_2014} for which TI was grown on the commercial InP substrates with minimal lattice mismatch and the highest sample quality. In the end, we compare the  values in samples B and D that both have 1 QL \BS~ seed layers. For sample D, we intentionally dope the \BST~with Cr, which can induce ferromagnetism at low temperature \cite{Jiang_2015}. At room temperature, the Cr doping mainly changes the transport properties and the SOC strength of the bulk states. The  values for samples B and D are \LIR $= 20 \pm 2$ pm and $22 \pm 3$ pm, respectively. Their similar spin-charge conversion efficiencies demonstrate that the properties of the TI bulk state do not play a significant role here, confirming the interface-dominated spin pumping phenomena. It is also important to note that values of \LIR~for samples B and D are lower than the value for YIG/\BS~(6QL). As in other studies of spin pumping into TIs, the interfacial condition presents a critical challenge for controlling the spin conversion efficiency \cite{Shiomi_2014,Deorani_2014,Jamali_2015}; in sample B, both YIG/\BS~and \BS~/\BST~interfaces will contribute to the formation of the surface states. Thus, structural defects and/or strain induced dislocations in the trilayer heterostructures can potentially result in the observed lower values. A thorough understanding about the correlation of the interfacial conditions of TI surfaces states and the spin-charge conversion efficiency requires further investigation.

Finally, we compare the spin transfer efficiency at YIG/\BS~to that at YIG/Pt. Note that Pt is an ideal spin sink and a well-studied non-magnetic material with large SOC \cite{Mosendz_2010,Wang_2014}. Figure 4(a) shows the inverse spin Hall spectrum of a YIG(30nm)/Pt(5nm) bilayer sample under 3 GHz and 100 mW microwave power when the $H$ field is in plane. The observed sign change of the spin pumping signal with field reversal is expected for the ISHE in a 3D material system \cite{Heinrich_2011,Mosendz_2010}. From the FMR linewidth broadening, the obtained YIG/Pt effective spin mixing conductance  is $(5.19 \pm 0.6) \times 10^{18} {\rm{m}}^{-2}$, which lies in the range of the values reported by other groups using spin pumping \cite{Hahn_2013,Wang_2014}. We compare this value with the obtained spin mixing conductance  at various \BS~thicknesses in Fig. 4(b). When \BS~is 6 QL thick, the spin mixing conductance at the YIG/\BS~interface is $(4.13 \pm 0.5) \times 10^{18} {\rm{m}}^{-2}$. Although there are some variations, the reported  values are in the range of $3 - 7 \times 10^{18} {\rm{m}}^{-2}$ when the \BS~thickness varies from 2 to 60 QL, which is essentially comparable to the determined value at the YIG/Pt interface, demonstrating an efficient spin transfer in YIG/TI heterostructures. It is important to note that in the large \BS~thickness regime, we do not observe an enhancement of \gsm, which is typically observed in the YIG/transition metal bilayers due to the decrease in backflow spin current caused by the spin diffusion in the bulk \cite{Tserkovnyak_2005,Jiao_2013}. This also confirms the TI surface states dominated spin-charge conversion mechanism.

In conclusion, we report robust spin pumping at room temperature in YIG/\BS~bilayers and other YIG/TI heterostructures. By measuring IREE voltages and interfacial spin current density, we determine the value of \LIR~and reveal its systematic behavior with \BS~thickness, demonstrating the dominant role of surface states in spin-charge conversion. The inferred IREE length indicates the important role of interface conditions in spin Hall physics in topological insulators. Further investigation is required for a thorough understanding of the correlation between the formation of the surface states and the variation of spin-charge conversion efficiency at the interfaces. 

\acknowledgements
The work at Penn State, Colorado State and University of Minnesota is supported by the Center for Spintronic Materials, Interfaces, and Novel Architectures (C-SPIN), a funded center of STARnet, a Semiconductor Research Corporation (SRC) program sponsored by MARCO and DARPA. NS and AR acknowledge additional support from ONR- N00014-15-1-2364. TL, HC, and MW acknowledge additional support from NSF-ECCS-1231598 and ARO-W911NF-14-1-0501. This work utilized (1) the College of Science and Engineering (CSE) Characterization Facility, University of Minnesota (UM), supported in part by NSF through the UMN MRSEC program (No. DMR-1420013); and (2) the CSE Minnesota Nano Center, UM, supported in part by NSF through the NNIN program.


\begin{thebibliography}{36}%
\bibitem{Kajiwara_2010} Y. Kajiwara, K. Harii, S. Takahashi, J. Ohe, K. Uchida, M. Mizuguchi, H. Umezawa, H. Kawai, K. Ando, K. Takanashi, S. Maekawa, and E. Saitoh, Nature {\bf 464}, 262 (2010).
\bibitem{Heinrich_2011} B. Heinrich, C. Burrowes, E. Montoya, B. Kardasz, E. Girt, Y.-Y. Song, Y. Y. Sun, and M. Z. Wu, Phys. Rev. Lett. {\bf 107}, 066604 (2011).
\bibitem{Mosendz_2010} O. Mosendz, V. Vlaminck, J. E. Pearson, F. Y. Fradin, G. E. W. Bauer, S. D. Bader, and A. Hoffmann, Phys. Rev. B {\bf 82}, 214403 (2010).
\bibitem{Du_2013} C. H. Du, H. L. Wang, Y. Pu, T. L. Meyer, P. M. Woodward, F. Y. Yang, and P. C. Hammel, Phys. Rev. Lett. {\bf 111}, 247202 (2013).
\bibitem{Hahn_2013} C. Hahn, G. de Loubens, O. Klein, M. Viret, V. V. Naletov, and J. Ben Youssef, Phys. Rev. B {\bf 87}, 174417 (2013).
\bibitem{Wang_2014} H. L. Wang, C. H. Du, Y. Pu, R. Adur, P. C. Hammel, and F. Y. Yang, Phys. Rev. Lett. {\bf 112}, 197201 (2014).
\bibitem{Sanchez_2013} J.-C. Rojas-Sánchez, L. Vila, G. Desfonds, S. Gambarelli, J. P. Attané, J. M. De Teresa, C. Magén, and A. Fert, Nat. Commun. {\bf 4}, 2944 (2013).
\bibitem{Shiomi_2014} Y. Shiomi, K. Nomura, Y. Kajiwara, K. Eto, M. Novak, K. Segawa, Y. Ando, and E. Saitoh, Phys. Rev. Lett. {\bf 113}, 196601 (2014).
\bibitem{Mellnik_2014} A. R. Mellnik, J. S. Lee, A. Richardella, J. L. Grab, P. J. Mintun, M. H. Fischer, A. Vaezi, A. Manchon, E.-A. Kim, N. Samarth, and D. C. Ralph, Nature {\bf 511}, 449 (2014).
\bibitem{Fan_2014} Y. Fan, P. Upadhyaya, X. Kou, M. Lang, S. Takei, Z. Wang, J. Tang, L. He, L.-T. Chang, M. Montazeri, G. Yu, W. Jiang, T. Nie, R. N. Schwartz, Y. Tserkovnyak, and K. L. Wang, Nat. Mater. {\bf 13}, 699 (2014).
\bibitem{Deorani_2014} P. Deorani, J. Son, K. Banerjee, N. Koirala, M. Brahlek, S. Oh, and H. Yang, Phys. Rev. B {\bf 90}, 094403 (2014).
\bibitem{Jamali_2015} M. Jamali, J. S. Lee, J. S. Jeong, F. Mahfouzi, Y. Lv, Z. Zhao, B. K. Nikolic, K. A. Mkhoyan, N. Samarth, and J.-P. Wang, Nano Lett. {\bf 15}, 7126 (2015).
\bibitem{Baker_2015}A. A. Baker, A. I. Figueroa, L. J. Collins-McIntyre, G. van der Laan, and T. Hesjeda, Scientific Reports {\bf 5}, 7907 (2015).
\bibitem{Wang_2015} Y. Wang, P. Deorani, K. Banerjee, N. Koirala, M. Brahlek, S. Oh, and H. Yang, Phys. Rev. Lett. {\bf 114}, 257202 (2015).
\bibitem{Kondou_arxiv} K. Kondou, R. Yoshimi, A. Tsukazaki, Y. Fukuma, J. Matsuno, K. S. Takahashi, M. Kawasaki, Y. Tokura, and Y. Otani, arXiv:1510.03572.
\bibitem{Hsieh_2009} D. Hsieh {\it et al.}, Nature {\bf 460}, 1101 (2009).
\bibitem{Hasan_RMP} M. Z. Hasan and C. L. Kane, Rev. Mod. Phys. {\bf 82}, 3045 (2010).
\bibitem{Qi2010} X. -L. Qi and S. -C. Zhang, Rev. Mod. Phys. {\bf 83,} 1057 (2011).
\bibitem{Chang_2014} H. Chang, P. Li, W. Zhang, T. Liu, A. Hoffmann, L. Deng, and M. Z. Wu, IEEE Magn. Lett. {\bf 5}, 6700104 (2014).
\bibitem{Kandala_2013} A. Kandala, A. Richardella, D. W. Rench, D. M. Zhang, T. C. Flanagan, and N. Samarth, Appl. Phys. Lett. {\bf 103}, 202409 (2013).
\bibitem{JSLee_2015} J. S. Lee, A. Richardella, D. Reifsnyder Hickey, K. A. Mkhoyan, and N. Samarth, Phys. Rev. B {\bf 92}, 155312 (2015).
\bibitem{Edelstein_1990} V. M. Edelstein, Solid State Commun. {\bf 73}, 225 (1990).
\bibitem{Shen_2014} K. Shen, G. Vignale, and R. Raimondi, Phys. Rev. Lett. {\bf 112}, 096601(2014).
\bibitem{Tserkovnyak_2005} Y. Tserkovnyak, A. Brataas, G. E.W. Bauer, and B. I. Halperin, Rev. Mod. Phys. {\bf 77}, 1375 (2005).
\bibitem{Liu_2015} L. Liu, A. Richardella, I. Garate, Y. Zhu, N. Samarth, and C.-T. Chen, Phys. Rev. B {\bf 91}, 235437 (2015).
\bibitem{Tang_2014} J. Tang, L.-T. Chang, X. Kou, K. Murata, E. S. Choi, M. Lang, Y. Fan, Y. Jiang, M. Montazeri, W. Jiang, Y. Wang, L. He, and K. L. Wang, Nano Lett. {\bf 14}, 5423 (2014).
\bibitem{Li_2014} C. H. Li, O. M. J. van't Erve, J. T. Robinson, Y. Liu, L. Li, and B. T. Jonker, Nat. Nanotechnol. {\bf 9}, 218 (2014).
\bibitem{YZhang_2010} Y. Zhang {\it et al.}, Nat. Phys. {\bf 6}, 584 (2010).
\bibitem{Neupane_2014} M. Neupane {\it et al.}, Nat. Commun. {\bf 5}, 3841 (2014).
\bibitem{Jiang_2015} Z. Jiang, C.-Z. Chang, C. Tang, P. Wei, J. S. Moodera, and J. Shi, Nano Lett. {\bf 15}, 5835 (2015).
\bibitem{Kandala_2015} A. Kandala, A. Richardella, S. Kempinger, C.-X. Liu, and N. Samarth, Nat. Commun. {\bf 6}, 7434 (2015).
\bibitem{Jiao_2013} H. J. Jiao and G. E. W. Bauer, Phys. Rev. Lett. {\bf 110}, 217602 (2013).

\end{thebibliography}
%

\clearpage

\begin{figure}
\includegraphics[width=120mm]{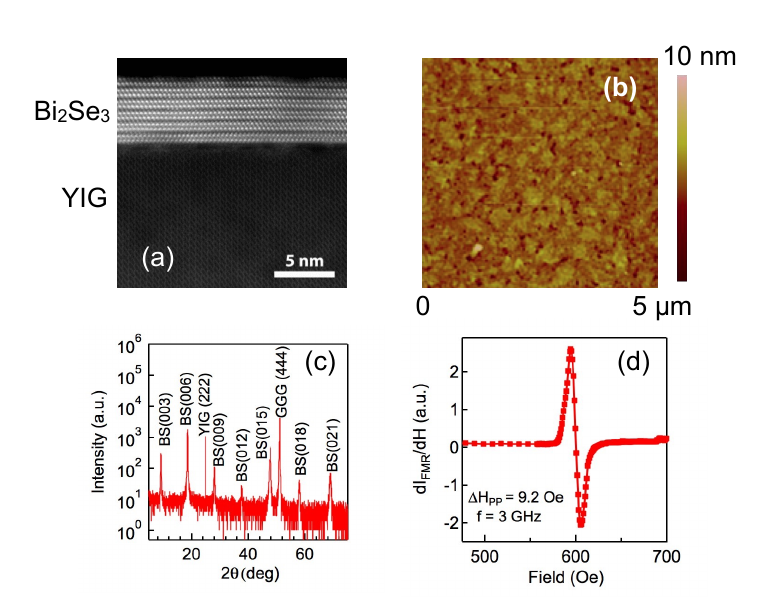} 
\caption{(Color online) (a) Cross-sectional high-angle annular dark-field scanning TEM image of the YIG/\BS~interface. (b) Atomic force microscopy  image of the YIG/\BS~ (6QL) sample with a surface roughness of 0.71 nm. (c) Semi-log $\theta - 2 \theta$ XRD scan of a YIG/\BS~(40QL) sample which exhibits clear x-ray scattering peaks from the (003) to (0021) planes of \BS. (d) A representative room-temperature FMR derivative spectrum of a 30-nm YIG film with an in-plane field, which gives a peak-to-peak line width of 9.2 Oe at $f = 3$GHz.}
\label{fig1}
\end{figure}
\newpage
\begin{figure}
\includegraphics[width=120mm]{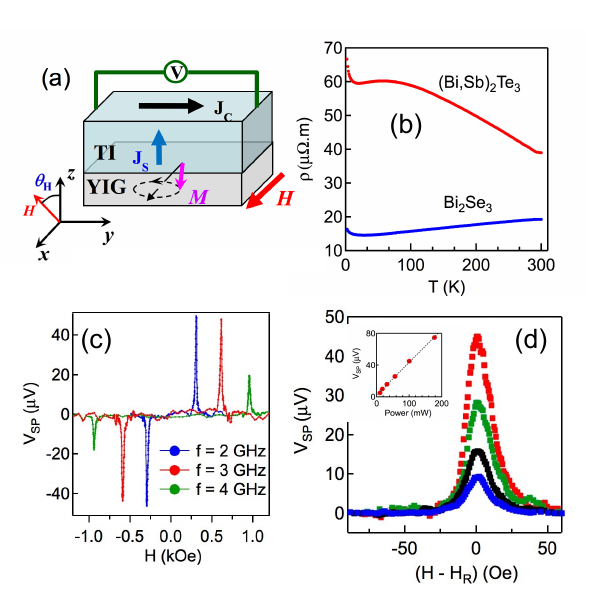} 
\caption{(Color online) (a) Schematic of the experimental setup for FMR spin pumping measurements. (b) Resistivity of 6 QL \BS~and \BST~thin films grown on YIG as a function of temperature. (c) $V_{SP}$ vs. $H$ spectra of YIG/\BS~(6QL) at $f =$ 2, 3, and 4 GHz using 100 mW microwave power. (d) $V_{SP}$ vs. $H$ spectra of the YIG/\BS~(6QL) sample for the microwave power of 18, 32, 56, and 100 mW at $f = 3$GHz. Inset: rf-power dependence of the corresponding $V_{SP}$ at $\theta_H = 90{^\circ}$.
}
\label{fig2}
\end{figure}

\newpage
\begin{figure*}
\includegraphics[width=120mm]{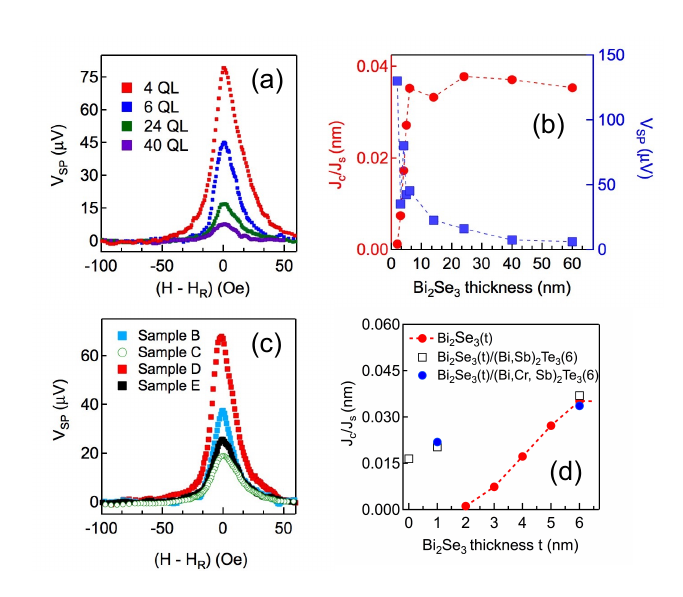} 
\caption{(Color online) (a) $V_{SP}$ vs. {\it H} spectra of YIG/\BS~(4QL), YIG/\BS~(6QL), YIG/\BS~(24QL), and YIG/\BS~(40QL) at $f = 3$ GHz using 100 mW microwave power. The $x$-axis is shifted by the resonance field ($H_R$) for clarity. (b) Dependence of $V_{SP}$~(blue points) and the spin-to-charge conversion efficiency $J_c / J_s$ determined by \LIR~(red points) on the \BS~thickness.  (c) $V_{SP}$ vs. $H$ spectra of control sample B: YIG/\BS(1QL)/\BST(6QL) (blue curve); sample C: YIG/\BS(6QL)/\BST(6QL) (green curve); sample D: YIG/\BS(1QL)/Cr$_{0.2}$(Bi$_{0.5}$Sb$_{0.5}$)$_{1.8}$Te$_3$(6QL) (red curve), and sample E: YIG/\BS(6 QL)/Cr$_{0.2}$(Bi$_{0.5}$Sb$_{0.5}$)$_{1.8}$Te$_3$(6QL) (black curve) at 3 GHz and 100 mW. (d) Comparison of $J_c/ J_s$ for the control samples with the corresponding values for YIG/\BS.}
\label{fig3}
\end{figure*}

\newpage
\begin{figure*}
\includegraphics[width=120mm]{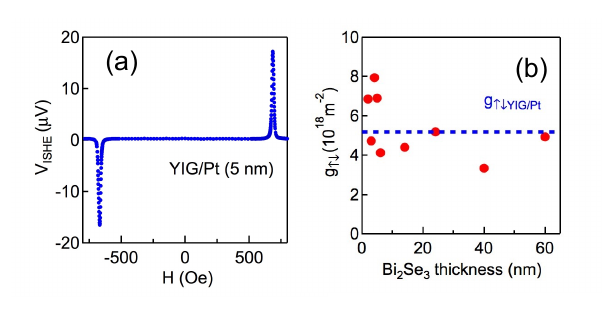} 
\caption{(Color online) (a) V$_{\rm{ISHE}}$ vs. $H$ spectra of YIG/Pt (5nm) bilayer at radio-frequency of 3 GHz and 100 mW microwave power. (b) Dependence of the YIG/\BS~interfacial spin mixing conductance \gsm~on \BS~thickness. The blue dashed line indicates the value of \gsm~at the YIG/Pt interface.
}
\label{fig4}
\end{figure*}

\end{document}